\documentstyle[twocolumn,aps,epsf]{revtex}

\def\i{{\rm i}}
\def\d{{\rm d}}
\def\e{{\rm e}}
\def\vector#1{{\bf #1}}
\def\vp{{\vector p}}
\def\vk{{\vector k}}
\def\vq{{\vector q}}

\def\vr{{\vector r}}

\def\dps{\displaystyle}

\def\Tc{{T_{\rm c}}}

\def\hightc{{high-$T_{\rm c}$ }}

\def\LSCO{\mbox{${\rm La_{2-{\it x}}Sr_{\it x}CuO_4}$}}
\def\BSCCO{\mbox{${\rm Bi_2Sr_2CaCu_2O_{8+\delta}}$}}

\def\TMTSFPF{\mbox{${\rm (TMTSF)_2PF_6}$}}

\def\SrRuO{\mbox{${\rm Sr_2RuO_4}$}}

\def\hsp#1{\hspace{#1ex}}

\def\Tc{{T_{\rm c}}}

\def\lsim{\stackrel{{\textstyle<}}{\raisebox{-.75ex}{$\sim$}}}
\def\gsim{\stackrel{{\textstyle>}}{\raisebox{-.75ex}{$\sim$}}}

\def\para{\parallel}

\def\kF{k_{{\rm F}}}

\def\omegaD{{\omega_{\rm D}}}

\def\qs{{q_{\rm s}}}

\begin{document}
\draft

\twocolumn[\hsize\textwidth\columnwidth\hsize\csname 
@twocolumnfalse\endcsname

\title{Anisotropic superconductivity mediated by phonons \\
       in layered compounds with weak screening effects}

\author{Hiroshi Shimahara and Mahito Kohmoto$^{*}$} 

\def\runtitle{Anisotropic superconductivity mediated by phonons \\
         in layered compounds with weak screening effect} 


\address{
Department of Quantum Matter Science, ADSM, Hiroshima University, 
Higashi-Hiroshima 739-8526, Japan \\
$^{*}$Institute for Solid State Physics, University of Tokyo, 
Kashiwa 277-8581, Japan
}

\date{Received ~~~ March 2001}

\maketitle

\begin{abstract}
Anisotropic pairing interactions mediated by phonons are examined 
in layer systems. 
It is shown that the screening effects become weaker 
when the layer spacing increases. 
Then the anisotropic components of the pairing interactions 
increase with the screening length 
since the momentum dependence changes. 
As a result, various types of anisotropic superconductivity occur 
depending on the parameter region. 
For example, $p$-wave superconductivity occurs when the short-range 
part of Coulomb repulsion is strong and the layer spacing is large. 
Two kinds of inter-layer pairing may occur 
when the layer spacing is not too large. 
Although the phonon contribution to the $d$-wave pairing interaction 
is weaker than the $p$-wave interaction, 
it increases with the layer spacing. 
Relevance of the present results to organic superconductors, 
\hightc cuprates, and \SrRuO \hsp{0.2} is discussed. 
\end{abstract}

\pacs{
}


]

\narrowtext

\begin{center}
{\bf I. INTRODUCTION}
\end{center}

Anisotropy of superconducting order parameter 
and the mechanism of pairing interactions 
in layered superconductors 
are recent subjects of much interest. 
In particular, \hightc cuprate superconductors, 
organic superconductors, 
and \SrRuO \hsp{0.25} compound have been studied by many authors.

There are some evidences that the order parameter has line nodes 
on Fermi-surface in \hightc cuprates. 
For example, an experiment and a theory on Josephson junction 
gave an evidence of ``$d$-wave'' order parameter 
in a cuprate superconductor~\cite{Tsu94,Sig92}. 
Linear temperature dependence of the penetration depth was observed 
at low temperatures~\cite{Har93}.

On the other hand, superconductivity in \SrRuO \hsp{0.25} 
is considered to be due to spin triplet pairing 
according to the results of a Knight shift measurement~\cite{Ish98} 
and a $\mu SR$ experiment~\cite{Luk98}. 
In the organic superconductor \TMTSFPF, a Knight shift measurement 
seems to support spin triplet pairing~\cite{Lee00,Koh00}.

On mechanism of the high transition temperature of the cuprates, 
pairing interactions of magnetic origin, 
such as exchange of spin fluctuations~\cite{Miy88,Shi88} 
and a super-exchange interaction between nearest neighbor spins, 
have been discussed by many authors, 
because of proximity to the antiferromagnetic phase. 
However, experimental results of isotope effect suggest 
that there are contributions to the superconductivity 
{from} phonon-mediated interactions in many \hightc 
cuprates~\cite{Kat87,Lea87,Bat87a,Fal87,Bat87b,Bat88,Fra94}. 
Absolute values of shifts of $\Tc$ 
are very large (0.2K $\sim$ 0.7K), 
but isotope effect exponents $\alpha$ are small 
because of the high transition temperature.

Abrikosov proposed a theory based on weak screening of 
Coulomb interactions and phonon-mediated pairing interactions 
in which anisotropic $s$-wave order parameter was obtained~\cite{Abr94}. 
In the presence of on-site Coulomb repulsion, 
extremely anisotropic $s$-wave order parameter 
with nodes was obtained~\cite{Abr95}. 
Bouvier and Bok also calculated an order parameter explicitly, 
and obtained anisotropic $s$-wave in the same model~\cite{Bou95}. 
Recently, it has been shown that $d$-wave superconductivity is 
reproduced in a similar model with 
antiferromagnetic fluctuations~\cite{Fri00,Cha00}.

We proposed in our previous paper~\cite{Shi01} 
that triplet pairing superconductivity can be induced 
by phonon-mediated interactions in ferromagnetic compounds, 
where singlet pairing is suppressed by Pauli paramagnetic effect.

The origin of the anisotropic components of pairing interactions 
mediated by phonons is briefly explained as follows. 
The screening effect limits electron-ion interactions within 
a range of the order of the screening length. 
Since the pairing interactions mediated by phonons are obtained 
by a second order perturbation of the electron-ion interactions, 
they also have a range of the same order. 
For example, the screening effect is taken into account 
as vertex corrections within diagramatic technique~\cite{Sch83}. 
When the screening length increases, 
the interactions are more localized in the momentum space. 
Hence, the anisotropic components of the interactions increase 
with the screening length.

In this paper, we examine layered superconductors 
with the phonon-mediated pairing interactions, 
extending it to systems with large layer spacing. 
The layered structure modifies the screening length 
and the pairing interactions significantly. 
It is shown that 
anisotropic components of the pairing interactions 
increase with the layer spacing. 
We argue from this result that some aspects of the layered 
superconductors, such as \SrRuO, organics, and cuprates, 
can be explained in the present model.

We also study an effect of anisotropy of density of states in 
square lattice systems. 
Although the effect of the anisotropy must be most remarkable 
when the Fermi-surface is near the van Hove singularities, 
we consider a system not necessarily near the van Hove singularity 
but a system with the density of states anisotropy within the layers. 
Enhancement of the transition temperature 
due to the van Hove singularities 
has been discussed by many authors. 
In particular, it has often been discussed 
that the effect is efficient especially 
for the $d_{x^2-y^2}$ pairing~\cite{Ohk87,Shi87,Shi88,Shi00a,Fri00,Cha00}. 
In this paper, 
we concentrate on the anisotropy 
but do not discuss the value of the density of states. 
It is shown that superconductivity is enhanced by the anisotropy 
for pairing of $d_{x^2-y^2}$ symmetry, 
but not for those of $d_{xy}$, $p_x$, and $p_y$ symmetries.

In section II, we define the model of the pairing interactions 
mediated by phonons. 
We derive expressions of 
the coupling constants for various types 
of anisotropic superconductivity. 
In section III, we examine the dependence on the layer spacing 
of the screening length and the pairing interactions. 
In section IV, we consider a situation in which inter-layer coupling 
is of the order of intra-layer coupling. 
In section V, we examine an effect of the anisotropy in the electron 
dispersion in square lattice systems. 
Section VI is devoted to discussion and summary.

\vspace{\baselineskip}
\begin{center}
{\bf II. SCREENING EFFECT AND PAIRING INTERACTIONS}
\end{center}

First, we introduce a model of pairing interactions. 
Abrikosov examined an effective pairing interaction mediated by phonons 
of the form 
\def\eqmodelAbr{(1)}
$$
     V(\vq) = g (\frac{\qs^2}{q^2 + \qs^2})^n 
              \frac{[\omega(\vq)]^2}
                   {(\xi_{\vk} - \xi_{\vk+\vq})^2 - [\omega(\vq)]^2} , 
     \eqno\eqmodelAbr
     $$
with $q = |\vq|$ and $\qs = l_{\rm s}^{-1}$, 
where $l_{\rm s}$ denotes the screening length~\cite{Abr94,Abr95}. 
A similar form corresponding to $n = 1$ is obtained by taking into 
account the screening effect in electron-phonon interactions 
as explained in a text book~\cite{Sch83}. 
If we put $n = 1$ for simplicity and $\xi_{\vk} - \xi_{\vk+\vq} = 0$ 
for the electrons near the Fermi-surface in eq.~{\eqmodelAbr} 
according to Abrikosov~\cite{Abr94}, we obtain a simplified form 
\def\eqmodel{(2)}
$$
     V(\vq) = - \frac{g \, \qs^2}{q^2 + \qs^2} . 
     \eqno\eqmodel
     $$
We examined this model in three dimensions 
in our previous paper~\cite{Shi01}.

In this paper, we consider the layer systems, 
and define lattice constants $a$ within the layers 
and $b$ between the layers. 
We take $x$ and $y$-axes in the direction of the lattice vectors 
within the layers, and $z$-axis perpendicular to the layers. 
In eq.~{\eqmodel}, the range of the interaction is 
$l_{\rm s}/b$ layers in the $z$-direction, 
while it is $l_{\rm s}/a$ sites in the layers. 
Thus, the anisotropy due to the layered structure is 
partially taken into account for the difference of the lattice constants 
$b$ and $a$. 
This model is qualitatively appropriate for long wave length 
such as $\lambda \sim q^{-1} \gg a, b, l_{\rm s}$. 
However, when $b \gg a$, 
the discrete layered structure in the inter-layer direction 
must be taken into account for shorter wave length 
$\lambda \sim q^{-1} \sim b$. 
Therefore we extend eq.~{\eqmodel} in the form 
\def\eqmodelex{(3)}
$$
     V(\vq) = - \frac{g \, \qs^2}{|\vq_{\para}|^2 + \qs^2} 
              - \frac{g' \, {q_{\rm s}'}^2}
                     {|\vq_{\para}|^2 + {q_{\rm s}'}^2}
                     \cos q_z b 
     \eqno\eqmodelex
     $$
for layer systems, 
where $\vq_{\para}$ is the momentum element in the layers. 
Here we have truncated the interaction at the nearest layers. 
The first and second terms correspond to the intra-layer and inter-layer 
interactions, respectively. 
The parameter $q_{\rm s}'$ is the inverse of the range of the interactions 
between electrons on the nearest layers in the $x$ and $y$-directions.

The gap equation of supercoductivity is written as 
\def\eqgapeq{(4)}
$$
     \Delta(\vk) = - \frac{1}{N} \sum_{\vk'} V(\vk - \vk')
                     W(\vk') \Delta(\vk') , 
     \eqno\eqgapeq
     $$
where 
\def\eqWdef{(5)}
$$
     W(\vk') = \frac{\tanh \frac{E(\vk')}{2T} }{2E(\vk')} 
     \eqno\eqWdef
     $$
with $E(\vk) = \sqrt{\epsilon_{\vk}^2 + [\Delta(\vk)]^2}$ 
and $N$ the number of lattice sites.

We put the gap function 
\def\eqgapkparakz{(6)}
$$
     \Delta(\vk) = \Delta_{\para}(\vk_{\para}) \, \eta(k_z) , 
     \eqno\eqgapkparakz
     $$
where $\vk_{\para} = (k_x,k_y)$ and 
$\eta(k_z)$ is a normalized function of the momentum component $k_z$. 
{From} eq.~{\eqmodelex}, 
the solution of the gap equation {\eqgapeq} at $T = \Tc$ has a form 
with $\eta(k_z) = 1$, $\sqrt{2} \cos k_z b$, or $\sqrt{2} \sin k_z b$. 
Then eq.~{\eqgapeq} is written as 
\def\eqgapeqparallel{(7)}
$$
     \Delta_{\para}(\vk_{\para}) 
       = - \frac{1}{N_{\para}} 
           \sum_{\vk'_{\para}} 
           V(\vk_{\para},\vk'_{\para}) W(\vk'_{\para}) 
           \Delta_{\para}(\vk'_{\para})  , 
     \eqno\eqgapeqparallel
     $$
where $N_{\para}$ denotes the number of sites in a layer, 
and $V(\vk_{\para},\vk'_{\para})$ denotes the averaged pairing 
interaction defined by 
\def\eqVavekzdef{(8)}
$$
     V(\vk_{\para},\vk'_{\para}) 
     \equiv \frac{b^2}{(2\pi)^2} 
            \int_{-\pi/b}^{\pi/b} \hsp{-2} \d k_z 
            \int_{-\pi/b}^{\pi/b} \hsp{-2} \d k_z' \, 
            \eta(k_z) \, V(\vk,\vk') \, \eta(k_z') . 
     \eqno\eqVavekzdef
     $$
Here we assume that the dispersion in the $z$-direction can be 
neglected in $\epsilon_{\vk}$ in the gap equation.

We consider cylindrically symmetic Fermi-surface from now on. 
Hence we put $|\vk_{\para}| = |\vk'_{\para}| = \kF$ 
in the pairing interactions eq.~{\eqVavekzdef} 
and obtain 
\def\eqVonFS{(9)}
$$
     V(\varphi - \varphi') \equiv 
     V(\vk_{\para},\vk'_{\para}) 
     = - \frac{g \, (\alpha - 1)}{\alpha - \cos (\varphi-\varphi')} , 
     \eqno\eqVonFS
     $$
with 
\def\eqalphadef{(10)}
$$
     \alpha = 1 + \frac{\qs^2}{2\kF^2} , 
     \eqno\eqalphadef
     $$
for $\eta(k_z) = 1$. 
On the other hand, for the order parameters with 
$\eta(k_z) = \sqrt{2} \cos k_z b$ and 
$\eta(k_z) = \sqrt{2} \sin k_z b$, 
the expression for $V(\varphi - \varphi')$ 
is obtained by replacing $g$ and $\alpha$ 
with $g'$ and $\alpha' = 1 + {q_{\rm s}'}^2/2\kF^2$, respectively, 
in eq.~{\eqVonFS}.

We expand the averaged interaction $V(\varphi - \varphi')$ as 
\def\eqVexpand{(11)}
$$
     \begin{array}{rcl}
     V(\varphi - \varphi') 
     & = & \dps{ 
          \sum_{m = 0}^{\infty} 
          V_m n_m \gamma_{m}(\varphi-\varphi') } \\[12pt]
     & = & \dps{ 
          \sum_{m = 0}^{\infty} 
          V_m (   \gamma_{m}(\varphi) \gamma_{m}(\varphi') 
                + {\bar \gamma}_{m}(\varphi) {\bar \gamma}_{m}(\varphi') ) , }
     \end{array}
     \eqno\eqVexpand
     $$
and the gap function 
$\Delta_{\para}(\varphi) = \Delta_{\para}(\vk_{\para})$ as 
\def\eqDeltaexpand{(12)}
$$
     \Delta_{\para}(\varphi)
        = 
          \sum_{m = 0}^{\infty} [
            \Delta_m  \gamma_{m}(\varphi) 
          + {\bar \Delta}_m {\bar \gamma}_{m}(\varphi) ] , 
     \eqno\eqDeltaexpand
     $$
where 
\def\eqgammamdef{(13)}
$$
     { \left \{ 
     \begin{array}{rcl} 
     \gamma_{m}(\varphi)         & = &  n_m \cos (m \varphi) \\[4pt]
     {\bar \gamma}_{m}(\varphi)  & = &  n_m \sin (m \varphi) , 
     \end{array}
     \right .}
     \eqno\eqgammamdef
     $$
with normalization factors 
\def\eqnmdef{(14)}
$$
     n_m = { \left \{ 
     \begin{array}{ll}
     1        & \mbox{~~for } m = 0 \\
     \sqrt{2} & \mbox{~~for } m \ne 0 . 
     \end{array}
     \right. }
     \eqno\eqnmdef
     $$
The expansion factor $V_m$ is calculated by 
\def\eqVmequation{(15)}
$$
     V_m = \frac{1}{n_m} \int_0^{2\pi} 
           \frac{\d \theta}{2\pi} \,
           \gamma_m(\theta)\, V(\theta) . 
     \eqno\eqVmequation
     $$
It is easy to perform the integration in eq.~{\eqVmequation}. 
For $\eta(k_z) = 1$, we obtain dimensionless coupling constants 
\def\eqlambdamresult{(16)}
$$
     \lambda_m = g N(0) \, 
                   \sqrt{ \frac{\alpha - 1}{\alpha + 1} } \, 
                   {\bigl [} \alpha - \sqrt{\alpha^2 - 1} 
                   {\bigr ]}^{m} . 
     \eqno\eqlambdamresult
     $$
Then the superconducting transition temperature $\Tc$ is obtained by 
\def\eqTc{(17)}
$$
     \Tc = 1.13 \, \omegaD \, \e^{-1/\lambda_{m}} , 
     \eqno\eqTc
     $$
with $ \lambda_{m} = - V_m N(0)$ {from} eq.~{\eqgapeq}, 
where $N(0)$ is the density of states per site of a given spin.

For $\eta(k_z) = \sqrt{2} \cos k_z b$ and $\sqrt{2} \sin k_z b$ 
we obtain a similar dimensionless coupling constant as 
\def\eqlambdamnnresult{(18)}
$$
     \lambda'_{m} = \frac{1}{2} g' N(0) \, 
                   \sqrt{ \frac{\alpha' - 1}{\alpha' + 1} } \, 
                   {\bigl [} \alpha' - \sqrt{{\alpha'}^2 - 1} 
                   {\bigr ]}^{m} , 
     \eqno\eqlambdamnnresult
     $$
for nearest neighbor layer pairings. 
The expression for $\Tc$ is the same as eq.~{\eqTc}.

Here, we note that a contribution 
{from} the short-range part of the Coulomb repulsion must be 
subtracted {from} $\lambda_0$ obtained above. 
For example in the tight binding model, 
the on-site Coulomb energy is estimated by 
\def\eqUexample{(19)}
$$
     U = \int \hsp{-1} \int \d^3 \vr \, \d^3 \vr' \, |w(\vr)|^2 \, 
     \frac{e^2}{4\pi \epsilon_0 |\vr-\vr'|} \, |w(\vr')|^2 , 
     \eqno\eqUexample
     $$
where $w(\vr)$ is the Wannier function. 
It is obvious that the energy $U$ is not included in 
our interaction energy eq.~{\eqmodelex}, 
since eq.~{\eqUexample} depends on the profile of the Wannier function. 
Equation~{\eqmodelex} describes the behaviors of pairing interactions 
of longer wave length, 
while the energy $U$ in eq.~{\eqUexample} is characterized 
by the local states of electrons on each lattice site.

Therefore we must consider the on-site Coulomb repulsion in addition to 
the pairing interaction of eq.~{\eqmodelex}. 
However, it reduces only the intra-layer $s$-wave pairing interaction 
but not the other anisotropic pairing interactions 
because of the symmetry. 
We define a parameter ${\tilde U}$ so that 
the $s$-wave interaction $\lambda_0$ 
is reduced by ${\tilde u} \equiv {\tilde U}N(0)$. 
The value of the parameter ${\tilde U}$ is not equal to $U$, 
because the retardation and spin fluctuation effects 
should be taken into account. 
We consider ${\tilde U}$ as a given parameter without estimating it 
microscopically.

\vspace{\baselineskip}
\begin{center}
{\bf III. DEPENDENCE ON THE LAYER SPACING OF 
THE ANISOTROPIC PAIRING INTERACTION}
\end{center}

In this section, 
we calculate anisotropic components of the effective pairing 
interactions as functions of the layer spacing $b$. 
We concentrate on the case of intra-layer pairing $\eta(k_z) = 1$ 
for a while.

The squared inverse of the screening length is 
\def\eqqsdef{(20)}
$$
     \qs^2 = \frac{e^2}{\epsilon_0} \rho(\mu) , 
     \eqno\eqqsdef
     $$
in Thomas-Fermi approximation, 
where $\rho(\mu)$ is total density of states of electrons 
per unit volume at chemical potential $\mu$. 
In layer systems, the total density of states per unit volume $\rho(\mu)$ 
is written in terms of the total density of states per unit area 
$\rho_{\para}^{\rm 2D}(\mu)$ in each layers as 
\def\eqrhoandrhotwodim{(21)}
$$
     \rho(\mu) = \rho_{\para}^{\rm 2D}(\mu)/b . 
     \eqno\eqrhoandrhotwodim
     $$
Here it is found that the screening becomes weaker 
when the layer spacing increases, 
because the volume density of electrons 
which contribute to screening decreases 
when the layer spacing increases. 
However, it should be noted that 
the screening length within a layer changes 
by the change of the inter-layer spacing $b$, 
even when the lattice constant $a$ in the layers is unchanged. 
Therefore the behavior of the screening length examined 
is not derived by a simple scale transformation 
in terms of $a$ and $b$ as the length scales.

We define a length scale $b_0$ as 
\def\eqbzerodef{(22)}
$$
     \alpha = 1 + \frac{\qs^2}{2 \kF^2} \equiv 1 + \frac{b_0}{b} , 
     \eqno\eqbzerodef
     $$
{from} eqs.~{\eqqsdef} and {\eqrhoandrhotwodim}. 
In a simple case, the length scale $b_0$ is estimated as follows. 
Assuming non-interacting two dimensional electron gas 
in $\rho^{(\rm 2D)}_{\para}(\mu)$, 
we obtain 
\def\eqbzeroinan{(23)}
$$
     b_0 = \frac{a^2}{\pi n a_{\rm H}} , 
     \eqno\eqbzeroinan
     $$
since $\rho^{(\rm 2D)}_{\para}(\mu) = m/\pi \hbar^2$ 
and $\kF a = \sqrt{2\pi n}$, 
where $n$ is the electron number per site. 
Here $a_{\rm H}$ denotes Bohr radius 
$a_{\rm H} = 4\pi\epsilon_0 \hbar^2/me^2 = 0.5292{\rm \AA}$. 
As an example, if $a \sim 4{\rm \AA}$ and $n \sim 1$ 
we have $b_0 \sim 9.6{\rm \AA}$ as a crude estimation.

Since the basic length scale $a_{\rm H}$ which is 
independent of the lattice constants $a$ and $b$ 
comes in eq.~{\eqbzeroinan}, 
changes not only of the ratio $b/a$ but also of the absolute values 
of $a$ and $b$ give rise to changes in the qualitative results.

\vspace{\baselineskip}
\begin{figure}[htb]
\begin{center}
\leavevmode \epsfxsize=6cm  
\epsfbox{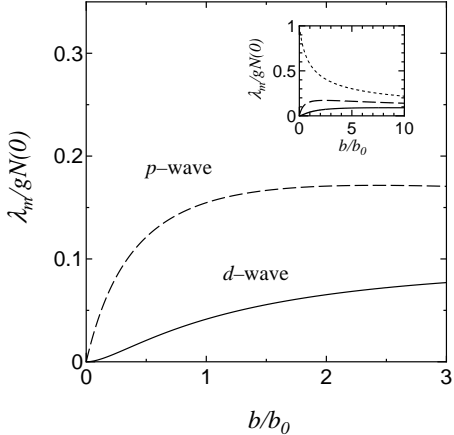}
\end{center}
\caption{
The dimensionless coupling constants $\lambda_m$ as a function of 
the layer spacing $b$. 
The solid and dashed lines show the results for $p$-wave ($m = 1$) 
and $d$-wave ($m = 2$), respectively. 
In the inset, 
the short dashed line shows the result for $s$-wave ($m = 0$). 
}
\label{fig:lambdam}
\end{figure}

\vspace{\baselineskip}
\begin{figure}[htb]
\begin{center}
\leavevmode \epsfxsize=6cm  
\epsfbox{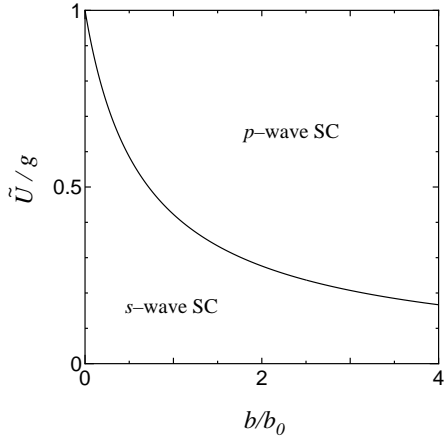}
\end{center}
\caption{
The phase diagram at $T = 0$ in $b$-${\tilde U}$ plane. 
SC stands for superconductivity. 
}
\label{fig:PDsp}
\end{figure}

Figure~\ref{fig:lambdam} shows the result of $\lambda_m$ 
as a function of the layer spacing $b$. 
It is seen that both $p$-wave and $d$-wave components of 
the pairing interactions increase with the layer spacing $b$. 
In particular, it is found that 
the $p$-wave components increase rapidly in the region $0 < b \lsim b_0$. 
As the inset shows, 
the $s$-wave component $\lambda_0/gN(0)$ is equal to 1 
in the limit of $b = 0$ and decreases with $b$. 
It remains larger than the other anisotropic components, 
but if the additional short-range Coulomb energy $U$ is sufficiently large 
so that $\lambda_0 - {\tilde u} < \lambda_1$, 
$p$-wave pairing occurs instead of $s$-wave pairing.

Figure~\ref{fig:PDsp} is the phase diagram at $T=0$ 
in $b$-${\tilde U}$ plane. 
It is found that $p$-wave superconductivity occurs in the region 
where the layer spacing $b$ is larger and the short-range repulsion 
expressed by ${\tilde U}$ is stronger. 
We will discuss the reality of such parameter values 
in the layered compounds in the last section.

On the other hand, for $d$-wave superconductivity to occur, 
some additional contribution to $\lambda_2$ 
or a negative contribution to $\lambda_1$ is needed, 
so that $\lambda_2$ becomes larger than $\lambda_1$. 
We examine an enhancement of $\lambda_2$ due to an anisotropy of 
the density of states later, 
and briefly discuss a contribution from the antiferromagnetic fluctuations 
in the last section.

\vspace{\baselineskip}
\begin{center}
{\bf IV. INTER-LAYER PAIRING}
\end{center}

In this section we consider a situation in which the inter-layer coupling 
constant $g'$ is of the same order as the intra-layer coupling constant $g$. 
The coupling constants would depend on the layer spacing $b$, 
but here we regard them as independent parameters. 
The condition $g' \sim g$ would actually be satisfied when $b$ is not 
too large. 
Then, we must consider the gap function of the form 
$\Delta(\vk) = \Delta_{\para}(\vk_{\para}) \eta(k_z)$ 
with $\eta(k_z) = \sqrt{2} \cos k_z b$ or $\sqrt{2} \sin k_z b$. 
The expansion of $\Delta_{\para}(\vk_{\para})$ by eq.~{\eqDeltaexpand} 
holds also in this case.

\begin{figure}[htb]
\begin{center}
\leavevmode \epsfxsize=6cm  
\epsfbox{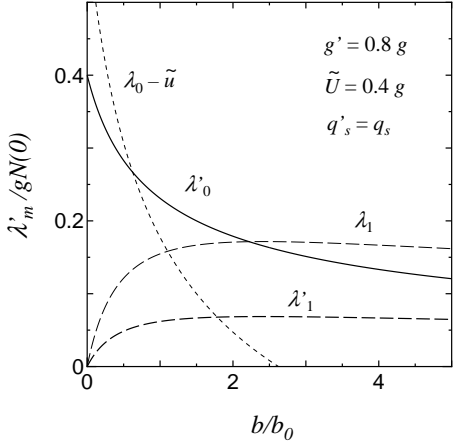}
\end{center}
\caption{
The dimensionless coupling constants $\lambda'_m$ 
of nearest neighbor layer pairing 
as a function of the layer spacing $b$. 
The thick solid and dashed lines show the results of the inter-layer 
pairing with $m = 0$ and $m = 1$, respectively, 
while the thin dashed and short dashed lines show the results of intra-layer 
pairing, $\lambda_1$ and $\lambda_0 - {\tilde u}$, respectively. 
}
\label{fig:nnlayerpairing}
\end{figure}

Figure~\ref{fig:nnlayerpairing} shows the dimensionless coupling 
constants $\lambda'_m$. 
A set of parameters, $g' = 0.8g$, ${\tilde U} = 0.4g$, 
and $q_{\rm s}' = \qs$ 
are taken as an example. 
For $b/b_0 \lsim 0.6$ and $b/b_0 \gsim 2.2$, 
intra-layer pairing (of $s$-wave and $p$-wave in each region, respectively) 
is favored. 
On the other hand, for $0.6 \lsim b/b_0 \lsim 2.2$, 
inter-layer pairing with $m = 0$ is favored. 
The gap function has a form such as 
\def\eqnnlayergap{(24)}
$$
     {\left \{ 
     \begin{array}{rcl}
     \Delta(\vk) & = & \Delta_{0} \sin k_z b \\
     \Delta(\vk) & = & \Delta_{0} \cos k_z b . 
     \end{array}
     \right .}
     \eqno\eqnnlayergap
     $$
The formar is an order parameter of triplet pairing, 
while the latter is that of singlet pairing. 
These gap functions have horizontal line nodes 
at $k_z = 0, \pm \pi/b$ and 
at $k_z = \pm \pi/2b$, respectively, 
but they are isotropic in the layers.

\vspace{\baselineskip}

\vspace{\baselineskip}
\begin{center}
{\bf V. EFFECT OF ANISOTROPY IN THE ELECTRON DISPERSION} 
\end{center}

In this section, we consider the square lattice systems, 
in which the electron dispersion depends on the direction of the momentum. 
We will show that 
the $d$-wave coupling constant $\lambda_2$ is enhanced 
for $d_{x^2-y^2}$ symmetry, but not for $d_{xy}$ symmetry, 
due to the anisotropy of the density of single-particle states. 
We define an angle-dependent density of states $\rho(\epsilon,\varphi)$ 
as a density of single-particle states per unit energy and unit angle.

In the square lattice system, 
the angle-dependent density of states at the Fermi-energy 
$\rho(0,\varphi)$ can be approximated by 
\def\eqrhophidef{(25)}
$$
     \rho(0,\varphi) \approx \rho_0 + \rho_4 \cos(4 \varphi) , 
     \eqno\eqrhophidef
     $$
where $\varphi$ is the angle between a momentum $\vp$ and $p_x$-axis. 
In addition, we regard $\kF$ as being constant, for simplicity.

\begin{figure}[htb]
\begin{center}
\leavevmode \epsfxsize=6cm  
\epsfbox{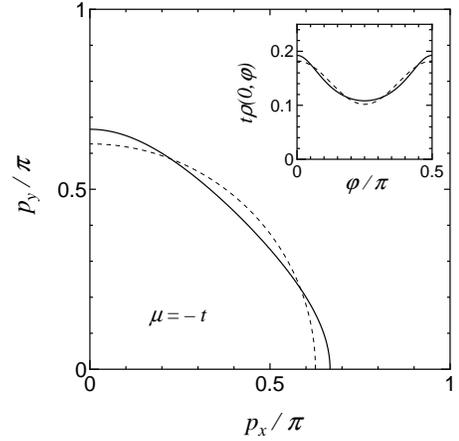}
\end{center}
\caption{
The Fermi-surface of a tight binding model with a chemical 
potential $\mu = - t$ (solid line), 
and the averaged isotropic Fermi-surface with $\kF \approx 1.97/a$ 
(short dashed line). 
The inset shows the angle-dependent density of states $\rho(0,\varphi)$ 
at the Fermi-energy. 
Short dashed line in the inset shows the behavior of $\rho(0,\varphi)$ 
approximated by eq.~{\eqrhophidef} 
with $\rho_0 = 0.142 t$ and $\rho_4 = 0.040 t$. 
}
\label{fig:FS}
\end{figure}

Figure~\ref{fig:FS} shows a verification of this simplified model 
in the square lattice tight binding model 
with a nearest neighbor hopping energy $t$ at $\mu = - t$. 
Although the Fermi-surface is nearly isotropic, 
the density of states $\rho(0,\varphi)$ varies 
with the direction $\varphi$. 
For example, when $\mu = -t$, 
$\rho_0 \approx 0.142$ and $\rho_4 \approx 0.040$ are estimated.

Regarding eq.~{\eqrhophidef} as an expansion of $\rho(\varphi,0)$, 
we could extend it into more general forms by adding terms 
$\rho_{4n} \cos(4n \varphi)$ with $n \geq 2$. 
Then the terms of $\rho_{4n}$ mix $\Delta_{m}$ of a small $m$ with 
$\Delta_{m'}$ of a large $m' = |m \pm 4n|$. 
However, since $V_m$ decreases rapidly with $m$ as seen by 
eq.~{\eqlambdamresult}, 
$\Delta_{m'}$s of such large $m'$ are small. 
Therefore the higher order terms in the expansion of $\rho(\varphi,0)$ 
can be omitted in practice.

In the gap equation, 
the anisotropic term proportional to $\rho_4 \cos 4 \varphi $ 
of the angle-dependent density of states 
mixes $\Delta_m \cos m\varphi$ with 
$\Delta_{|m-4|}\cos[(m-4)\varphi]$ 
and $\Delta_{m+4} \cos[(m+4)\varphi]$, 
but it does not affect equations for ${\bar \Delta}_m \sin m \varphi$. 
Therefore we only consider equations for $\Delta_m$. 
For general $m$, we can write the gap equation at $T = \Tc$ as 
\def\eqgaprhophi{(26)}
$$
     \begin{array}{rcl}
     \Delta_m & = & \dps{
        - \lambda_m^{(0)} \log \frac{2\e^{\gamma} \omegaD}{\pi \Tc}
                    }\\[12pt]
              &   & \dps{
                \times 
                {\Bigl [} \Delta_m 
                  + \frac{\rho_4}{2 \rho_0} 
                {\bigl \{}
                    \frac{n_m}{n_{m+4}} \Delta_{m+4} 
                  + \frac{n_m}{n_{|m-4|}} \Delta_{|m-4|} 
                {\bigr \}}
                {\Bigr ]} , 
                    }
     \end{array}
     \eqno\eqgaprhophi
     $$
where we define 
$\lambda_m^{(0)} \equiv V_m \rho_0 = V_m N(0)$ is 
the dimensionless coupling constant for the isotropic case.

\vspace{\baselineskip}

Since $V_3$, $V_4$, $V_5$ $\cdots$ are much smaller than $V_0$ and $V_1$, 
the terms proportional to $\rho_4$ can be neglected 
in eq.~{\eqgaprhophi} for $m = 0$ and 1. 
Hence, $\lambda_0$ and $\lambda_1$ are not modified by $\rho_4$. 
On the other hand, for $m = 2$, we cannot omit the term of 
$\Delta_{|m-4|}$ in eq.~{\eqgaprhophi} since $|m - 4| = 2$. 
Neglecting the term of $\Delta_{m+4} = \Delta_6$ because $V_6 \ll V_2$, 
we obtain 
\def\eqgaprhophidwave{(27)}
$$
     \begin{array}{rcl}
     \Delta_2 & = & \dps{
                \lambda_2^{(0)} 
                \, {\Bigl [} 
                      1 + \frac{\rho_4}{2 \rho_0} 
                   {\Bigr ]}
                \, \log \frac{2\e^{\gamma} \omegaD}{\pi \Tc}
                \, \Delta_2 }\\[8pt]
              & \equiv & \dps{ 
                \lambda_2 
                \, \log \frac{2\e^{\gamma} \omegaD}{\pi \Tc}
                \, \Delta_2 , 
                }
     \end{array}
     \eqno\eqgaprhophidwave
     $$
where we define an effective coupling constant 
$\lambda_2 \equiv \lambda_2^{(0)} (1 + \rho_4/2\rho_0)$, 
which gives $\Tc$ by eq.~{\eqTc}.

Therefore, it is found that 
$d_{x^2-y^2}$-wave pairing is favored more than 
$d_{xy}$-wave pairing 
by the enhancement factor $(1 + \rho_4/2\rho_0)$, 
since $\rho_4$ changes the equation for $\Delta_m$ 
but does not for ${\bar \Delta}_m$ as mentioned above. 
The enhancement factor $1 + \rho_4/2 \rho_0$ 
is estimated to be 1.14 for $\mu = - t$, 
and 1.22 for $\mu= -0.5 t$. 
On the competition with $p$-wave pairing, 
those values are not large enough to change the sign of 
$\lambda_2 - \lambda_1$. 
Therefore, another non-phonon contribution seems to be needed for 
$d$-wave pairing to occur.

\begin{figure}[htb]
\begin{center}
\leavevmode \epsfxsize=6cm  
\epsfbox{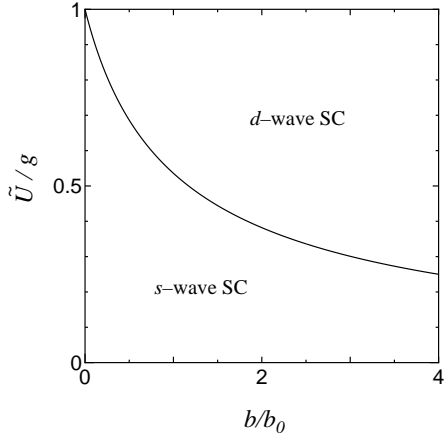}
\end{center}
\caption{
The phase diagram of $s$-wave superconductivity 
and $d_{x^2-y^2}$-wave superconductivity at $T = 0$ 
in $b$-${\tilde U}$ plane 
in the absence of $p$-wave superconductivity. 
SC stands for superconductivity. 
}
\label{fig:PDsd}
\end{figure}

Figure~\ref{fig:PDsd} are phase diagram in the absence of 
$p$-wave superconductivity. 
It is found that $d$-wave superconductivity is favored 
when the layer spacing $b$ is larger 
and the short-range Coulomb repulsion ${\tilde U}$ is stronger. 
If there are additional contributions to $d$-wave superconductivity 
mentioned above, the phase boundary is shifted downward.

\vspace{\baselineskip}

\vspace{\baselineskip}
\begin{center}
{\bf VI. DISCUSSION AND SUMMARY}
\end{center}

We have examined pairing interactions mediated by phonons 
in the layer systems. 
The screening of Coulomb interactions becomes weaker 
when the layer spacing $b$ increases. 
Then anisotropic components of the pairing interactions 
increase with the layer spacing $b$ 
since the momentum dependence of the interactions changes. 
In particular, $p$-wave superconductivity occurs 
for large $b$ and strong short-range Coulomb repulsion ${\tilde U}$, 
even in the absence of any additional non-phonon interactions.

It was found that the $p$-wave coupling constant $\lambda_1$ increases 
rapidly with the layer spacing $b$ in the region $b \lsim b_0$, 
where $b_0$ is a length scale defined by eq.~{\eqbzerodef}. 
For the rapid increase of $\lambda_1$, 
the condition $\lambda_0 - {\tilde u} < \lambda_1$ 
is realized more easily in layer systems 
than in usual three dimensional systems, 
where $\lambda_0$ denotes the $s$-wave coupling constant and 
$-{\tilde u}$ is a negative contribution to $s$-wave pairing 
due to the short-range Coulomb repulsion 
discussed near eq.~{\eqUexample}. 
Hence triplet pairing superconductivity is favored 
in layered compounds. 

We have also examined inter-layer pairing. 
In some region of the parameter space, 
for example $0.6 \lsim b/b_0 \lsim 2.2$ 
for the parameters indicated in Fig.~\ref{fig:nnlayerpairing}, 
the gap function has horizontal line nodes parallel to the layers. 
In this case, 
the solutions of singlet pairing and triplet pairing 
of eq.~{\eqnnlayergap} degenerate. 
If some effect due to spin fluctuations, ferromagnetic correlations, 
or spin-orbit coupling removes this degeneracy, 
inter-layer triplet pairing may occur.

In \SrRuO \hsp{0.25} compounds, 
existence of the line nodes was supported by some experiments 
such as temperature dependences of specific heat and NMR relaxation 
rate~\cite{Mae00}. 
However, direction of the line nodes does not seem clear at the present. 
Line nodes vertical to the layers were indicated 
by ultrasound attenuation~\cite{Lup00}, 
whereas almost isotropic state was indicated 
by thermal conductivity~\cite{Iza01}. 
The isotropic state can be consistent with the specific heat and 
NMR experiments, if the horizontal line nodes are assumed~\cite{Shi98}.

In the present theory, inter-layer triplet pairing 
reproduces the horizontal line nodes, 
while the intra-layer triplet pairing is a candidate 
for the vertical line nodes. 
For the latter pairing, 
we need some additional mechanism 
for the vertical line nodes to occur, 
because isotropic states such as $p_x + \i p_y$ have the lowest free 
energy in the present isotropic system. 
Consistent explanation of the experimental results within the 
present theory remains for a future study.

In order to discuss the reality of the phonon-mediated anisotropic 
superconductivity, 
we crudely estimate the parameters for 
the \SrRuO \hsp{0.2} compound 
and quasi-one-dimensional organic superconductors 
from the observed transition temperature $\Tc \sim 1.5{\rm K}$. 
We assume triplet pairing here, 
although for the organics it might be rather controversial. 
The results of the parameter values do not strongly depend 
on the direction of the line nodes.

Roughly speaking, $b \gsim b_0$ is satisfied in both kinds of compounds. 
Thus we have $\lambda_1 \sim 0.15 \times gN(0)$ 
for intra-layer triplet pairing 
from Fig.~\ref{fig:lambdam}, 
while $\lambda_0' \sim 0.22 \times gN(0)$ 
for inter-layer pairing 
from Fig.~\ref{fig:nnlayerpairing}. 
On the other hand, if we assume $\omegaD \sim 1000{\rm K}$ and 
$\Tc \sim 1.5{\rm K}$, we have $\lambda_1 \sim 0.151$ 
(or $\lambda_0' \sim 0.151$). 
Therefore, we obtain $gN(0) \sim 1.0$ and 0.69, respectively. 
For such choices of parameter values, 
it is likely that $s$-wave pairing is suppressed. 
For example, for intra-layer pairing, 
since we obtained $gN(0) \sim 1$ above, 
the $s$-wave coupling constant is estimated as 
$\lambda_0 \approx 0.58 \times gN(0) \sim 0.58$ at $b = b_0$ 
by eq.~{\eqlambdamresult}. 
Thus the difference of $\lambda_0$ and $\lambda_1$ is about 0.42, 
which corresponds to $V_0 - V_1 \sim 0.42/N(0) \sim 0.42 W$, 
where $W$ is the band width. 
Therefore $s$-wave pairing is suppressed when 
the magnitude of the negative contribution ${\tilde U}$ 
due to the short-range Coulomb repulsion 
is larger than $0.42 W \sim W/2$. 
Although this estimation is crude, 
the value $\sim W/2$ seems realistic as the order of the magnitude.

On the other hand, for $d$-wave superconductivity to occur 
in the present model, 
it is needed that $p$-wave pairing is suppressed 
for some extra reason as well as $s$-wave pairing, 
or that there are some additional contributions to $d$-wave pairing. 
For this problem, 
we examined effect of the anisotropy of the electron dispersion. 
It was found that the $d$-wave coupling constant $\lambda_2$ 
is enhanced by the anisotropy for $d_{x^2-y^2}$ symmetry, 
while not for $d_{xy}$ symmetry and $p_x$, $p_y$ symmetries. 
However, the enhancement does not seem to be large enough to realize 
the $d$-wave superconductivity.

This might suggest an existence of a non-phonon contribution to 
the $d$-wave pairing interaction in the cuprates. 
For example, 
many authors discussed that 
antiferromagnetic spin fluctuations may contribute to $d$-wave 
pairing interactions~\cite{Eme86,Bea86,Miy88,Shi88,Shi89,Fri00,Cha00}. 
In particular, such interactions enhance $d$-wave pairing efficiently 
in the presence of the van Hove singularities~\cite{Shi88} 
in the square lattice systems.

However, 
even if we assume that 
a non-phonon contribution is indispensable for high-$\Tc$, 
the present theory suggests that there is a large phonon contribution 
to the $d$-wave pairing interactions especially in layer systems 
for the weak screening. 
This result is consistent with the observed large shifts of $\Tc$ 
as absolute values due to the 
isotope effect~\cite{Kat87,Lea87,Bat87a,Fal87,Bat87b,Bat88,Fra94}.

It was also found that the coupling constant $\lambda_2$ increases 
with the layer spacing $b$. 
This behavior might be a reason why 
the transition temperature of \BSCCO \hsp{0.25} 
is much higher than that of \LSCO. 
Since $\Tc$ is a sensitive function of $\lambda_m$, 
such a slight enhancement of $\lambda_2$ may increase $\Tc$ 
considerably.

In conclusion, phonon-mediated pairing interactions include 
anisotropic components of various symmetries. 
In particular, they are large when the screening effect is weak 
due to the layered structure of the system. 
As a result, the interactions induce various types of anisotropic 
superconductivity depending on values of the energy parameters 
and lattice constants. 
In particular, triplet superconductivity is favored 
for large layer spacing and strong short-range repulsion. 
For anisotropic superconductivity, $\Tc$ increases with the layer 
spacing $b$. 
The phonon-mediated pairing interactions may play some 
essential roles in the anisotropic superconductivity of 
layered compounds, such as \SrRuO, the organic superconductors, 
and the \hightc cuprates.

\begin{center}
ACKNOWLEDGEMENTS
\end{center}

The authors wish to thank Prof.~J.~Friedel for useful discussions. 
A part of this work was supported by a grant for Core Research 
for Evolutionary Science and Technology (CREST) from Japan Science 
and Technology Corporation (JST).




\begin{thebibliography}{99}
\bibitem{Tsu94}
  C.~C.~Tsuei, J.~R.~Kirtley, C.~C.~Chi, L.~S.~Y.-Jahnes, A.~Gupta, 
  T.~Shaw, J.~Z.~Sun, and M.~B.~Ketchen, 
  Phys. Rev. Lett. {\bf 73}, 593 (1994). 
\bibitem{Sig92}
  M.~Sigrist and T.~M.~Rice, J. Phys. Soc. Jpn. {\bf 61}, 4283 (1992). 
\bibitem{Har93}
  W.~N.~Hardy, D.~A.~Bonn, D.~C.~Morgan, R.~Liang, and K.~Zhang, 
  Phys. Rev. Lett. {\bf 70}, 3999 (1993); 
  D.~A.~Bonn, R.~Liang, T.~M.~Riseman, D.~J.~Baar, D.~C.~Morgan, 
  K.~Zhang, P.~Dosanjh, T.~L.~Duty, A.~MacFarlane, 
  G.~D.~Morris, J.~H.~Brewer, and W.~N.~Hardy, 
  Phys. Rev. B {\bf 47}, 11314 (1993). 
\bibitem{Ish98}
  K.~Ishida, H.~Mukuda, Y.~Kitaoka, K.~Asayama, Z.~Q.~Mao, 
  Y.~Mori, Y.~Maeno, Nature {\bf 396}, 658 (1998). 
\bibitem{Luk98}
  G.~M.~Luke Y.~Fudamoto, K.~M.~Kojima, M.~I.~Larkin, J.~Merrin, 
  B.~Nachumi, Y.~J.~Uemura, Y.~Maeno, Z.~Q.~Mao, Y.~Mori, H.~Nakamura, 
  M.~Sigrist, 
  Nature {\bf 394}, 558 (1998). 
\bibitem{Lee00}
  I.~J.~Lee, D.~S.~Chow, W.~G.~Clark, J.~Strouse, 
  M.~J.~Naughton, P.~M.~Chaikin, S.~E.~Brown, cond-mat/0001332. 
\bibitem{Koh00}
  M.~Kohmoto and M.~Sato, cond-mat/0003211. 
\bibitem{Miy88}
  K.~Miyake, T.~Matsuura, K.~Sano, and Y.~Nagaoka: 
  J. Phys. Soc. Jpn. {\bf 57} (1988) 722. 
\bibitem{Shi88}
  H.~Shimahara and S.~Takada: J. Phys. Soc. Jpn. {\bf 57} (1988) 1044. 
\bibitem{Kat87}
  H. Katayama-Yoshida, T. Hirooka, A. J. Mascarenhas, Y. Okabe, 
  T. Takahashi, T. Sasaki, A. Ochiai. T. Suzuki, J. I. Pankove, 
  T. Ciszek, S. K. Deb, Jpn. J. Appl. Phys. {\bf 26}, L2085 (1987). 
\bibitem{Lea87}
  K. J. Leary, H. C. zur Loye, S. W. Kelter, T. A. Faltens, 
  W. K. Ham, J. N. Michaels, A. M. Stacy, 
  Phys. Rev. Lett. {\bf 59}, 1236 (1987). 
\bibitem{Bat87a}
  B. Batlogg, G. Kourouklis, W. Weber, R. J. Cava, A. Jayaraman, 
  A. E. White, K. T. Short, L. W. Rupp, E. A. Rietman, 
  Phys. Rev. Lett. {\bf 59}, 912 (1987). 
\bibitem{Fal87}
  T. A. Faltens, W. K. Ham, S. W. Keller, K. J. Leary, J. N. Michaels, 
  A. M. Stacy, H. C. zur Loye, T. W. Barbee, III, L. C. Bourne, 
  M. L. Cohen, S. Hoen, A. Zettl, 
  Phys. Rev. Lett. {\bf 59}, 915 (1987). 
\bibitem{Bat87b}
  B. Batlogg, R. J. Cava, A. Jayaraman, R. B. van Dover, 
  G. A. Kourouklis, S. Sunshine, D. W. Murphy, L. W. Rupp, H. S. Chen, 
  A. White, K. T. Short, A. M. Mujsce, E. A. Rietman, 
  Phys. Rev. Lett. {\bf 58}, 2333 (1987). 
\bibitem{Bat88}
  B. Batlogg, R. J. Cava, M. Stavola, 
  Phys. Rev. Lett. {\bf 60}, 754 (1988). 
\bibitem{Fra94}
  J. P. Franck, in {\it Physical Properties of High Temperature 
  Superconductors} IV, ed. D. M. Ginsberg (World Scientific 
  Singapore, 1994) p. 189. 
\bibitem{Abr94}
  A. A. Abrikosov, 
  Physica C {\bf 222}, 191 (1994). 
\bibitem{Abr95}
  A. A. Abrikosov, 
  Physica C {\bf 244}, 243 (1995). 
\bibitem{Bou95}
  J. Bouvier and J. Bok, Physica C {\bf 249} 117, (1995). 
\bibitem{Fri00}
  J. Friedel and M. Kohmoto, to be published. 
\bibitem{Cha00}
  I. Chang, J. Friedel, and M. Kohmoto, 
  Europhys. Lett. {\bf 50}, 782 (2000). 
\bibitem{Shi01}
  H.~Shimahara and M.~Kohmoto, cond-mat/0011188. 
\bibitem{Sch83}
  J.~R.~Schrieffer, {\it Theory of superconductivity}, 
  pp.152 (Addison-Weslay, New York, 1983). 
\bibitem{Ohk87}
  F.~J.~Ohkawa, 
  Jpn. J. Appl. Phys. {\bf 26}, L652 (1987); 
  J. Phys. Soc. Jpn. {\bf 56}, 2267 (1987). 
\bibitem{Shi87}
  H.~Shimahara and S.~Takada, Jpn. J. Appl. Phys. {\bf 26} L1674 (1987). 
\bibitem{Shi00a}
  H.~Shimahara, Y.~Hasegawa, and M.~Kohmoto, 
  cond-mat/9910349; J. Phys. Soc. Jpn. {\bf 69}, 1598 (2000). 
\bibitem{Mae00}
  Y.~Maeno, 
  Physica B {\bf 281-282}, 865 (2000), and references therein. 
\bibitem{Lup00}
  C.~Lupien, W.~A.~MacFarlane, C.~Proust, and L.~Tailefer, 
  Z.~Q.~Mao, and Y.~Maeno, 
  cond-mat/0101319, to be pulished in Phys. Rev. Lett. 
\bibitem{Iza01}
  K.~Izawa, H.~Takahashi, H.~Yamaguchi, Y.~Matsuda, M.~Suzuki, 
  T.~Sasaki, T.~Fukase, Y.~Yoshida, R.~Settai, Y.~Onuki, 
  cond-mat/0012137, to be published in Phys. Rev. Lett. 
\bibitem{Shi98}
  It has been discussed by many authors. 
  {[} H.~Shimahara, unpublished, 
  presentation 27aYA-10 in JPS autumn meeting (1998, Okinawa); 
  Y.~Hasegawa, K.~Machida, and M.~Ozaki, 
  J. Phys. Soc. Jpn. {\bf 69}, 336 (2000); 
  M.~E.~Zhitomirsky and T.~M.~Rice, 
  cond-mat/0102390.{]}
\bibitem{Eme86}
  V.~J.~Emery: Synth. Met. {\bf 13} (1986) 21.
\bibitem{Bea86}
  M.~T.~Beal-Monod, C.~Bourbonnais, and V.~J.~Emery: 
  Phys. Rev. B {\bf 34} (1986) 7716. 
\bibitem{Shi89}
  H.~Shimahara, J. Phys. Soc. Jpn. {\bf 58} (1989) 1735; 
  H.~Shimahara, {\it Proceeding of the Physics and Chemistry of
  Organic Superconductors}, edited by G. Saito and S. Kagoshima
  (Springer-Verlag, Berlin, Heidelberg, New York, 1990), p.73. 
\end{thebibliography}
\end{document}